\documentclass[twocolumn, preprintnumbers,amsmath,amssymb,balancelastpage]{revtex4}
\pdfoutput=1

\usepackage{putexPRL}
\usepackage{graphicx}
\usepackage{dsfont}
\usepackage{amsmath}
\usepackage{array}
\usepackage{epstopdf}
\usepackage{enumerate}
\usepackage{youngtab}
\usepackage{tensor}
\usepackage{braket}
\usepackage[normalem]{ulem}
\usepackage{slashed}
\usepackage[aligntableaux=center]{ytableau}
\usepackage[utf8]{inputenc}
\usepackage[
      colorlinks=true,
      linkcolor=blue,
      urlcolor=blue,
      filecolor=black,
      citecolor=red,
      ]{hyperref}
\usepackage{braket}
\usepackage{mathtools}
\usepackage{rotating}
\usepackage{tabularx}
\newcolumntype{Y}{>{\centering\arraybackslash}X}
\newcolumntype{C}[1]{>{\centering\arraybackslash}p{#1}}
\usepackage{todonotes}
\usepackage{xcolor}
\definecolor{LightCyan}{rgb}{0.7,1,1}
\definecolor{Gray}{gray}{0.9}
\usepackage{xspace}

\newcommand{\abs}[1]{\left\lvert #1 \right\rvert}

\newcommand {\be} {\begin {equation}}
\newcommand {\ee} {\end {equation}}

\newcommand {\bes} {\begin {equation*}}
\newcommand {\ees} {\end {equation*}}

\newcommand{\es}[2] {\begin{equation} \label{#1} \begin{split} #2 \end{split} \end{equation}}

\newcommand{\cO}{{\mathcal O}}

\def\le{\left}
\def\ri{\right}

\newcommand{\beq}{\begin{equation}}
\newcommand{\eeq}{\end{equation}}

\newcommand{\ov}{\over}

\def\ie{\begin{equation}\begin{aligned}}
\def\fe{\end{aligned}\end{equation}}

\def\<{\langle}
\def\>{\rangle}

\def\beg{\begin{equation}\begin{gathered}}
\def\eeg{\end{gathered}\end{equation}}

\def\bea{\begin{equation}\begin{aligned}}
\def\eea{\end{aligned}\end{equation}}

\begin{document}

\title{Evidence for 3d bosonization from monopole operators}

\author{Shai M.~Chester}         
\affiliation{
Jefferson Physical Laboratory, Harvard University, Cambridge, MA 02138, USA\\
Center of Mathematical Sciences and Applications, Harvard University, Cambridge, MA 02138, USA
		}
\author{\'Eric Dupuis}
\affiliation{D\'epartement de physique, Universit\'e de Montr\'eal, Montr\'eal (Qu\'ebec), H3C 3J7, Canada}

\author{William Witczak-Krempa}
\affiliation{D\'epartement de physique, Universit\'e de Montr\'eal, Montr\'eal (QC), H3C 3J7, Canada}
\affiliation{Centre de Recherches Math\'ematiques, Universit\'e de Montr\'eal; Montr\'eal (QC), H3C 3J7, Canada}
\affiliation{Institut Courtois, Universit\'e de Montr\'eal, Montr\'eal (QC), H2V 0B3, Canada}

\begin{abstract}
We give evidence for 3d bosonization in Conformal Field Theories (CFTs) by computing monopole operator scaling dimensions in 2+1 dimensional quantum electrodynamics (QED3) with Chern-Simons level $k$ and $N$ complex bosons in a large $N,k$ expansion. We first consider the $k=0$ case, where we show that scaling dimensions previously computed to subleading order in $1/N$ can be extrapolated to $N=1$ and matched to $O(2)$ Wilson-Fisher CFT scaling dimensions with around 5\% error, which is evidence for particle-vortex duality. We then generalize the subleading calculation to large $N,k$ and fixed $k/N$, extrapolate to $N=k=1$, and consider monopole operators that are conjectured to be dual to non-degenerate scalar operators in a theory of a single Dirac fermion. We find matches typically with 1\% error or less, which is strong evidence of this so-called `seed' duality that implies a web of 3d bosonization dualities among CFTs. 
\end{abstract}

\maketitle
\nopagebreak

{\bf Introduction---}IR duality is when two quantum field theories that are completely different at short distances (the UV), nonetheless flow to the same conformal field theory (CFT) at long distances (the IR). While duality is common in two spacetime dimensions, in 3d this phenomenon is much more rare. For many years, the only experimentally relevant example in 3d was particle/vortex duality \cite{Peskin:1977kp,PhysRevLett.47.1556}, which conjectures that the $O(2)$ Wilson-Fisher fixed point is dual to 3d quantum electrodynamics (QED3) with $N=1$ complex bosonic field and $k=0$ Chern-Simons level, the so-called abelian Higgs model. Recently, new dualities were conjectured between QED3 with various Chern-Simons levels and matter content \cite{PhysRevX.5.031027,PhysRevB.48.13749,PhysRevB.89.235116}, and these dualities were shown to be part of a so-called web of dualities that generically relates CFTs with fermionic matter to bosonic matter \cite{Karch:2016sxi,Seiberg:2016gmd}, and is thus an example of 3d bosonization. This duality web can be derived from a conjectured `seed' duality, which relates QED3 with $N=1$ boson and $k=1$ to the free theory of a single complex two-component fermion. The central idea is that the Chern-Simons term effectively attaches flux (a magnetic instanton or monopole) to the boson, leading to an additional Berry phase and Fermi statistics for the boson-flux composite.

QED3 at finite $N$ and $k$ is strongly coupled in the IR, which makes it hard to verify these dualities. While the conjectured dualities satisfy kinematic consistency checks such as 't Hooft anomalies  \cite{Seiberg:2016gmd}, we would ideally like to check local dynamical observables such as critical exponents (i.e.\/ scaling dimensions). When $k=0$, the theory can be modeled on the lattice \cite{Kajantie:2004vy,Karthik:2018rcg}, which found evidence for particle/vortex duality by comparing the lowest scaling dimensions $\Delta^O_q$ from $O(2)$ lattice studies \cite{Hasenbusch:2019jkj,Banerjee:2017fcx} of operators with charge $q$ under the $U(1)\cong O(2)$ global symmetry, to lattice estimates $\Delta^Q_q$ of the dual operator scaling dimensions in QED3 \footnote{The $O(2)$ lattice studies in fact have more precision than shown here, as we show later in Table \ref{pvtab}.}:
\es{oldevidence}{
\Delta^O_0&=1.511,\quad \Delta^O_{\frac12}=.5191,\quad \Delta^O_{1}=1.236,\quad \Delta^O_{\frac32}=2.109,\\
\Delta^Q_0&=1.508,\quad \Delta^Q_{\frac12}=.48,\quad\quad\Delta^Q_{1}=1.23,\quad\;\; \Delta^Q_{\frac32}=2.15.
}
Lattice methods have more difficulty when $k\neq0$ due to the sign problem, however, so it has been difficult to numerically verify the seed duality for 3d bosonization. Instead, the duality has been motivated by an uncontrolled flow \cite{Gur-Ari:2015pca,Kachru:2016rui} from more well established supersymmetric dualities \cite{Giveon:2008zn}, as well as as an extrapolation to $N_C=1$ of 3d bosonization for quantum chromodynamics (QCD3) at large colors $N_C$ and $k$ \cite{Aharony:2015mjs}, which has been checked at leading order in $1/N_C$ starting with \cite{Aharony:2012nh,Aharony:2011jz,Giombi:2011kc}.

Here we give evidence for both particle/vortex duality, and the seed bosonization duality by computing the scaling dimension of monopole operators in scalar QED3, and matching these to the dimensions of the operators in the dual theories. We will do this by considering QED3 in the limit of large $N$ scalars, and also large $k$ and fixed $\kappa\equiv k/N$ for the bosonization case, where monopole operator scaling dimensions can be computed in a $1/N$ expansion \cite{Murthy:1989ps,Borokhov:2002ib}. For $k=0$, the scaling dimensions have already been computed to subleading order in \cite{Dyer:2015zha}, so we will simply extrapolate these results to $N=1$ and compare to scaling dimensions of the critical $O(2)$ model as computed from the conformal bootstrap \cite{Chester:2019ifh,Liu:2020tpf} and lattice \cite{Banerjee:2017fcx}. For nonzero $k$, we will extend the leading order calculation in \cite{Chester:2017vdh} to sub-leading order for general $\kappa$, extrapolate to $N=\kappa=1$, and compare to scaling dimensions of non-degenerate scalar operators in the free fermion theory \footnote{The subleading calculation was previously computed at large $N,k$ for the specific case of $\kappa=1$ for the lowest charged monopole \cite{Chester:2021drl}, but this is dual to a free fermion operator, not a scalar, and so did not match the conjectured duality.}. In all cases, we find that our perturbative calculation matches the conjectured dualities with a relative error of just a few percent, as shown in Tables \ref{pvtab} and \ref{ktab}.

The rest of this letter is organized as follows.  We first introduce monopole operators and discuss our new calculation of their scaling dimension at large $N,k$. We then review how these operators are expected to map to the dual theories, and compare our new results. We end with a discussion of our results. Technical details are discussed in the Appendices.

{\bf Monopoles at large $N,k$---}
Monopole operators are defined in three dimensional Abelian gauge theories as local operators that are charged under the topological global symmetry $U(1)_\text{top}$ \cite{Polyakov:1975rs, Borokhov:2002ib}, whose conserved current and charge are
\es{topDef}{
j_\text{top}^\mu=\frac{1}{8\pi}\epsilon^{\mu\nu\rho}F_{\nu\rho}\,, \qquad q=\frac{1}{4\pi}\int_{\Sigma}F \,,
}
where $F_{\nu \rho} \equiv \partial_\nu A_\rho - \partial_\rho A_\nu$ is the gauge field strength with spacetime index $\mu=1,2,3$, $\Sigma$ is a closed two-dimensional surface, and $j_\text{top}^\mu$ is conserved due to the Bianchi identity. In the normalization \eqref{topDef}, the charge $q$ is restricted by Dirac quantization to take the values $q\in\mathbb{Z}/2$. As in \cite{Chester:2017vdh, Borokhov:2002ib, Pufu:2013vpa, Dyer:2013fja, Chester:2015wao, Metlitski:2008dw, Dyer:2015zha, Borokhov:2002cg,Dupuis:2019uhs,Dupuis:2019xdo,Dupuis:2021yej,Dupuis:2021flq}, we will compute the scaling dimension of the lowest dimension monopole operators using the state-operator correspondence, which identifies the scaling dimensions of monopole operators of charge $q$ with the energies of states in the Hilbert space on $S^2\times\mathbb{R}$ with $4\pi q$ magnetic flux through the sphere \cite{Borokhov:2002ib}. The ground state energy on $S^2\times\mathbb{R}$ can then be computed in the large $N$ and $k$ limit using a saddle point expansion. When $k\neq0$, the Chern-Simons term induces a gauge charge proportional to $q$, so that the naive $S^2\times \mathbb{R}$ vacuum must be dressed by charged matter modes. Following \cite{Chester:2017vdh}, we can enforce this dressing by computing the small temperature $T\equiv\beta^{-1}$ limit of the thermal free energy on $S^2\times S^1_\beta$, where the saddle point value of the holonomy of the gauge field on $S^1_\beta$ acts like a chemical potential for the matter fields. This dressing will make the monopole transform in a nontrivial representation under the $SU(N)$ flavor symmetry with a nonzero spin for the $SO(3)$ rotation symmetry.

We begin by writing the conformally invariant action of QED3 with $N$ complex scalars $\phi^i$ on $S^2\times S^1_\beta$ as \footnote{QED3 was shown to flow to an IR CFT at large $N$ \cite{Appelquist:1988sr}, and this behavior is believed to persist at finite $N$ except perhaps for $N=2$ and $k=0$.} 
\es{action2}{
\mathcal{S}&=\int\! d^3x\Big[  { \sqrt{g}}  \left(\abs{(\nabla_\mu-iA_\mu)\phi^i}^2+(\frac{1}{4}+i\lambda)|\phi^i|^2\right)\\
&\qquad\qquad\qquad\qquad\qquad\qquad-\frac{ik}{4\pi}\epsilon^{\mu\nu\rho}A_\mu\partial_\nu A_\rho\Big]\,,
}
where $g$ is the determinant of the metric, $\lambda$ is a Hubbard-Stratonovich field, and $i=1,\dots, N$.  We are interested in computing the thermal free energy ${F}_{q,\kappa}$ in the presence of a magnetic flux $\int dA=4\pi q$ through $S^2$. We can integrate out the matter fields in the path integral on this background to get
\es{free}{
\hspace{-.1in}e^{-\beta F_{q,\kappa}}\hspace{-.025in}=&\hspace{-.05in}\int \hspace{-.05in}DA\exp\hspace{-.05in}\left[-N\tr\log\hspace{-.025in}\big[\frac14+i\lambda-(\nabla_\mu-iA_\mu)^2\big]\right.\\
&\left.+iN\int d^3x\left(\frac{\kappa}{4\pi}\epsilon^{\mu\nu\rho}A_\mu\partial_\nu A_\rho+{\sqrt{g}}   \lambda\right)\right]\,,
}
where $\kappa= k/N$. We now expand $A_\mu$ and $\lambda$ around a saddle point by taking
\es{saddle}{
A_\mu=\mathcal{A}_\mu+a_\mu\,,\qquad i\lambda=\mu+i\sigma\,,
}
where $a_\mu$ and $\sigma$ are fluctuations around a background $A_\mu=\mathcal{A}_\mu$ and $i\lambda=\mu$ that satisfy
\es{saddleCond}{
\frac{\delta F_{q,\kappa}[A_\mu,\lambda]}{\delta A_\mu}\Bigg\vert_{\sigma=a_\mu=0}=\frac{\delta F_{q,\kappa}[A_\mu,\lambda]}{\delta \lambda}\Bigg\vert_{\sigma=a_\mu=0}=0\,.
}
On $S^2\times S^1_\beta$ with magnetic flux $4\pi q$, the most general such background is $\mu$ constant and $\mathcal{A}_\mu^q$ given by
\es{background}{
\mathcal{A}_\tau=-i\alpha\,,\qquad \mathcal{F}_{\theta\phi}d\theta\wedge d\phi=q\sin\theta d\theta\wedge d\phi\,, 
}
where $\alpha=i\beta^{-1}\int_{S^1_\beta}A$ is a real constant called the holonomy of the gauge field.

Since the integrand in \eqref{free} is proportional to $N$, the thermal free energy $F_{q,\kappa}$ can then be expanded at large $N$ as
\es{largeN}{
F_{q,\kappa}=NF_{q,\kappa}^{(0)}+F_{q,\kappa}^{(1)}+\frac1NF_{q,\kappa}^{(2)}+\dots\,,
}
where $ F_{q,\kappa}^{(0)}$ comes from evaluating $F_{q,\kappa}$ at the saddle point and $ F_{q,\kappa}^{(1)}$  comes from the functional determinant of the quantum fluctuations around the saddle point. The scaling dimension $\Delta_{q,\kappa}$ is then obtained from the zero temperature limit as
\es{largeN}{
\Delta_{q,\kappa}=N\Delta_{q,\kappa}^{(0)}+\Delta_{q,\kappa}^{(1)}+\dots\,, \quad \Delta_{q,\kappa}^{(n)}\equiv \lim_{\beta\to\infty}F_{q,\kappa}^{(n)}\,.
}
At leading order there are many degenerate monopoles when $k\neq0$, due to the different ways of dressing the bare monopole, which can be detected from the $O(\beta^{-1})$ terms in $F_{q,\kappa}^{(0)}$. This leads to degeneracy breaking contributions to $\Delta_{q,\kappa}^{(1)}$ that were shown \cite{Chester:2017vdh} to depend on spin, but whose explicit form was not worked out in general. 
 
 The leading order $F_{q,\kappa}^{(0)}$ was computed in \cite{Chester:2017vdh} for general $k,q$ by fixing $\mu$ and $\alpha$ from the saddle point equations \eqref{saddleCond}, setting $A_\mu$ and $\lambda$ in \eqref{free} to their saddle point values \eqref{background}, and then doing the resulting mode expansion. We review the details of this calculation in Appendix \ref{sec:details}, and give some of the resulting $\Delta_{q,\kappa}^{(0)}$ in Tables \ref{pvtab} and \ref{ktab}. The sub-leading $F_{q,\kappa}^{(1)}$ comes from  expanding \eqref{free} to quadratic order in the fluctuations $a_\mu$ and $\sigma$ around the saddle point values to get the Gaussian integral
\es{free1}{
&\exp(-\beta F_{q,\kappa}^{(1)})=\hspace{-.025in}\int \hspace{-.025in}DaD\sigma\exp\Big[-\frac{N}{2}\int d^3xd^3x'\\
&\times\sqrt{g}\sqrt{g'}\big(a_\mu(x)K^{\mu\nu}_q(x,x')a_\nu(x')\\
&+\sigma(x)K^{\sigma\sigma}_q(x,x')\sigma(x')+2\sigma(x)K^{\sigma\nu}_q(x,x')a_\nu(x')\big)\Big]\,,
}
where the kernels are expressed by expectation values of the matter fields for the saddle point values of $A_\mu$ and $\lambda$. These kernels can be computed in terms of the thermal Green's function 
$\langle\phi^i(x)\phi^*_j(x')\rangle=\delta^i_{j}G^q(x,x')$,
which was computed for general $q,\kappa$ in \cite{Chester:2017vdh}. We give explicit expressions in Appendix \ref{sec:details}, where we explain how to use these kernels to compute the small temperature expansion of $F_{q,\kappa}^{(1)}$, which yields the sub-leading $\Delta_{q,\kappa}^{(1)}$. For $\kappa=0$, this calculation was performed in \cite{Dyer:2015zha,DeLaFuente:2018uee}, and we list some of their results in Table \ref{pvtab}.  For $\kappa\neq0$, we have additional parity breaking contributions to the matter kernels, which makes the calculation much more challenging. When $\kappa=1$ and $q=1/2$, it was shown in \cite{Chester:2021drl} that the Green's function simplifies, so that matter kernels could be computed in closed form and used to compute the scaling dimension. 
In Appendix \ref{sec:details}, we extend this calculation to general $q,\kappa$ using an algorithmic approach, which yields the scaling dimensions in Table \ref{ktab}.

\begin{table}
\centering
\begin{tabular}{c||c|c|c|c|c}
$q$ & $ \Delta_{q,0}^{(0)} $ &  $\Delta_{q,0}^{(1)}$ &  $N=1$ & $O(2)$ & Error (\%)  \\ 
\hline
\hline
$1/2$&0.12459 & 0.38147& 0.50609 &{\Blue 0.519130434}&2.5\\
$1$&0.31110 &0.87452&1.1856&{\Blue 1.23648971}&4.1\\
$3/2$&0.54407 &1.4646&2.0087&{\Blue2.1086(3)}&4.7\\
$2$&0.81579 &2.1388&2.9546&{\Blue3.11535(73)}&5.2 \\
$5/2$&1.1214 &2.8879&4.0093&{\Red4.265(6)}&5.8 \\
$3$&1.4575 &3.7053&5.1628&{\Red5.509(7)}&6.3 \\
$7/2$&1.8217 &4.5857&6.4074&{\Red6.841(8)}&6.3 \\
$4$&2.2118 &5.5249&7.7367&{\Red8.278(9)}&6.5 \\
$9/2$&2.6263 &6.5194&9.1458&{\Red9.796(9)}&6.6 \\
$5$&3.0638 &7.5665&10.630&{\Red11.399(10)}&6.7 \\
\hline
\end{tabular}
\caption{Scaling dimensions $\Delta_{q,0}=N\Delta_{q,0}^{(0)}+\Delta_{q,0}^{(1)}+O(1/N)$ for charge $q$ scalar monopole operators in QED3 with $N$ scalars and $k=0$ in a large $N$ expansion~\cite{Dyer:2015zha,DeLaFuente:2018uee} extrapolated to $N=1$, compared to values of the dual operators in the critical $O(2)$ model as computed from the {\Blue numerical bootstrap ($q\leq 2$)} and {\Red lattice ($q>2$)}, along with the relative errors from the comparison.}
\label{pvtab}
\end{table} 
{\bf Duality Comparison---}
We will now extrapolate the large $N$ monopole scaling dimensions to $N=1$ and compare to the conjectured dual theories. For $k=0$, we expect the monopole operators of charge $q$ to be dual to the lowest dimension scalar operators of charge $q$ in the critical $O(2)$ Wilson-Fisher CFT, where $U(1)_\text{top}$ is identified with the $O(2)$ flavor symmetry. The scaling dimensions of operators with $q=1/2,1,3/2,2$ have been determined using the conformal bootstrap \cite{Chester:2019ifh,Liu:2020tpf}, while higher values of $q$ were determined with less accuracy using lattice methods \cite{Hasenbusch:2019jkj,Banerjee:2017fcx}. We compare these values in Table \ref{pvtab} \footnote{The large $N$ results are shown to 5 significant digits for $\kappa=0$ and 4 digits for $\kappa=1$, but any number of digits can be computed in principle. The bootstrap results for $q=1/2,1$ are rigorous \cite{Chester:2019ifh}, while for $q=3/2,2$ they were non-rigorously extracted from the spectrum and so have an uncertainty \cite{Liu:2020tpf}. The  lattice results in \cite{Hasenbusch:2019jkj,Banerjee:2017fcx} are less precise, and we give their uncertainty.}, and find that the monopole scaling dimensions match their expected duals with just a few percent relative error, which gradually grows with $q$. It is remarkable that the contribution of the quantum correction $\Delta_{q,0}^{(1)}$ exceeds the leading
saddle-point one by a factor of more than 2. This extends the previous lattice evidence for the singlet \cite{Kajantie:2004vy} and $q=1/2,1,3/2$ \cite{Karthik:2018rcg} monopole scaling dimensions as reviewed in \eqref{oldevidence}. Monopoles with $q=1/2$ and higher $N$ were also successfully matched to lattice calculations in antiferromagnets with SU($N$) symmetry that can be described by an effective CP$^{N-1}$ gauge theory as in Eq.~(\ref{action2}) with $k=0$~\cite{Dyer:2015zha}, so the sub-leading computation seems accurate for general $N$. Note that all our large-$N$ estimates are strictly below the estimates from other methods, which is also true for the boson-fermion duality that we now discuss. 

\begin{table}
\centering
\begin{tabular}{c||c|c|c|c|c}
$q$ & $ \Delta_{q,1}^{(0)} $ &  $\Delta_{q,1}^{(1)}$ &  $N=1$ & Fermion & Error (\%)   \\ 
\hline
\hline
$1/2$ & 1 & $-0.2789$ & 0.7211 & 1 & 28\\
 ${\Purple 1}$ & ${\Purple 2.5833}$ & {\Purple $-0.6312$} & {\Purple 1.952} &${\Purple2}$&${\Purple 2.4 }$\\
$3/2$&4.5873 & $-1.052$ & 3.535 & 4& 15 \\
$2$&6.9380 & $-1.534$ & 5.404 & 6& 9.9 \\
$5/2$&9.5904 & $-2.070$ & 7.521 & 8& 6.0 \\
${\Purple3}$&${\Purple12.514}$ & {\Purple $-2.655$} & {\Purple 9.859 } &${\Purple 10}$&${\Purple 1.4}$\\
${\Purple6}$&${\Purple34.727}$ & {\Purple $-7.032$} & {\Purple 27.70} &${\Purple 28}$&${\Purple 1.1}$\\
${\Purple10}$&${\Purple74.141}$ & {\Purple $-14.71$} & {\Purple 59.43} &${\Purple 60}$&${\Purple 0.95}$\\
${\Purple15}$&${\Purple135.67}$ & {\Purple $-26.64$} & {\Purple 109.03} &${\Purple 110}$&${\Purple 0.88}$\\
${\Purple21}$&${\Purple224.23}$ & {\Purple $-43.76$} & {\Purple 180.5} &${\Purple 182}$&${\Purple 0.84}$\\
\hline
\end{tabular}
\caption{Scaling dimensions $\Delta_{q,1}=N\Delta_{q,1}^{(0)}+\Delta_{q,1}^{(1)}+O(1/N)$ for charge $q$  monopole operators in QED3 with $N$ scalars and $k/N=1$ in a large $N,k$ expansion extrapolated to $N=k=1$, compared to values of the dual operators in the free fermion CFT, along with the relative errors from the comparison. We expect the comparison to be most precise when the operator is a unique scalar $q=1,3,6,\dots$, as denoted in {\Purple purple}.}
\label{ktab}
\end{table} 

We next consider the extrapolation of the large $N,k$ monopole scaling dimensions to $N=k=1$, where the theory is conjectured to be dual to a single free fermion $\psi_\alpha$ with spinor index $\alpha=1,2$. Monopoles of charge $q$ should be dual to the lowest dimension operator formed by $2q$ fermions, where we identify $U(1)_\text{top}$ with the $U(1)$ flavor symmetry of the complex fermion \footnote{For this duality, the parity symmetry that is manifest in the free fermion comes from the duality of $k=1$ QED3 to $k=-1$ \cite{Seiberg:2016gmd}. The monopole scaling dimension calculation only depends on $|k|$, so it automatically respects this parity duality.}. Due to the antisymmetry of the fermions and the equations of motion, the lowest operator must sometimes include derivatives, and so there will be multiple such operator with different spins for different contractions of the indices. For instance, while the lowest $q=1/2$ operator is the spin half $\psi_\alpha$, and the lowest $q=1$ is the spin zero $\psi_{[\alpha}\psi_{\beta]}$, already at $q=2$ the lowest dimension operators are $\psi_{\alpha_1} \psi_{\alpha_2} \slashed{\partial}_{\alpha_3\alpha_4}\psi_{\alpha_5}\slashed\partial_{\alpha_6\alpha_7}\psi_{\alpha_8}$, where the two ways of contracting the indices give spin zero or two. We can count the lowest dimension operators of a given $q$ by expanding the $S^2\times S$ partition function for the free fermion in characters of primary operators following \cite{Dolan:2005wy}, which we do in Appendix \ref{sec:count}. These operators are unique scalars when \cite{Komargodski:2021zzy}
\es{uniqueq}{
q=n (n + 1)/2\,,\qquad n=1,2,3,\dots\,,
}
in which case the scaling dimension is
\es{uniquedel}{
\Delta^\text{ferm}_q=\frac23 q\sqrt{8q+1} \,,
}
which corresponds to the energy of $n$ completely filled energy shells on $S^2\times\mathbb{R}$. We compare the monopole scaling dimensions to the dual free fermion operators in Table \ref{ktab}. We find a match with relative error of typically 1\% or less for values of $q$ in \eqref{uniqueq} when the operator is a unique scalar. For other values we do not find such a precise match, presumably because there are other contributions to the monopole scaling dimension in this case, such as the spin-dependent degeneracy breaking terms discuss above. As $q$ increase, the match improves for all $q$, especially for the unique scalar $q$.

{\bf Discussion---}
In this work we considered the scaling dimensions of monopoles in QED3 with $N$ scalars and Chern-Simons level $k$ as computed to sub-leading order at large $N,k$. When $k=0$, this computation was performed in \cite{Dyer:2015zha}, which we extended to the case of general $q$ and $\kappa\equiv k/N$. When $k=0$ and $N=1$ the theory is dual to the $O(2)$ Wilson-Fisher theory, while when $k=N=1$ the theory is dual to a single free fermion. We found evidence of each conjectured duality by extrapolating the large $N,k$ results, and found matches with just a few percent relative error in each case. For the $k=0$ case, all the monopoles are unique scalars and we found good matches for all $q$, which extends the evidence of particle/vortex duality beyond the lowest few $q$ considered in previous lattice studies \cite{Kajantie:2004vy,Karthik:2018rcg}. For the $k=1$ case, the monopoles are only unique scalars for certain $q$, which is where we found matches to good precision. This match is the first quantitative evidence for 3d bosonization, and the duality we consider in fact implies a large web of other dualities as discussed in \cite{Seiberg:2016gmd,Karch:2016sxi}.

Looking ahead, we would like to understand better why the large $N$ calculation is less accurate when the dual operator is degenerate or has nonzero spin. It was observed in \cite{Chester:2017vdh} that monopoles in scalar QED3 are degenerate at leading large $N$, which leads to spin-dependent degeneracy breaking contributions at sub-leading order. Unfortunately, we do not know how to compute this contribution for general $q,\kappa$, and even for $q=1/2$ and $\kappa=1$ where this contribution was computed in \cite{Chester:2021drl}, it did not improve the results. In fact, this contribution was found to be negative in this case, which implies that some temperature dependent terms in the free energy must be complex, which suggests that the saddle we chose was unstable in this case. It is thus possible that a different saddle point might be required when the monopole has spin. 

On the other hand, we have observed a curious coincidence that if we take the scaling dimension $\Delta_q^\text{ferm}$ in the free fermion theory for $q$ in \eqref{uniqueq} when the operator is a unique scalar and analytically continue to all $q\in\mathbb{Z}/2$, then this matches to high precision to all our estimates for the monopole scaling dimension, not just $q$ in \eqref{uniqueq} as before. For instance, for the lowest few $q$ that are not unique scalars we get the comparison
with \eqref{uniquedel}:
\es{magic}{
\Delta_{1/2}^\text{ferm}&=.7454,\; \Delta_{3/2}^\text{ferm}=3.606,\; \Delta_{2}^\text{ferm}=5.498,\\
\Delta_{1/2}^\text{mono}&=.7211,\; \Delta_{3/2}^\text{mono}=3.535,\; \Delta_{2}^\text{mono}=5.404,
}
where $q=1/2$ and $q=3/2$ in the free fermion picture are unique operators of spin $1/2$ and $3/2$, respectively, while for $q=2$ there are two degenerate lowest dimension operators of spins zero and 2 as shown in \eqref{listOps}. This analytic continuation of $q$ could be explained in the large $q$ expansion \cite{Hellerman:2015nra,Cuomo:2021qws}, where the effective action that describes unique scalar operators must be corrected to describe more general operators \cite{Komargodski:2021zzy}. Perhaps a similar correction should be added to the large $N$ monopole calculation for non-scalar or degenerate monopoles.

We would also like to understand better why the large $N$ calculation of monopole scaling dimensions works so well when computed to just sub-leading order \footnote{The large $N$ expansion also works well for some sphere free energies in 3d \cite{Klebanov:2011td}.}. For other operators in scalar QED3 that are constructed from matter fields, such as the lowest dimension singlet computed for $k=0$ in \cite{PhysRevLett.32.292}, the large $N$ expansion seems much less accurate when compared against lattice estimates such as \cite{Kajantie:2004vy}. Perhaps the large charge expansion \cite{Hellerman:2015nra,Cuomo:2021qws} could also explain this, as this expansion was shown to work well for the scaling dimension of monopole operators \cite{DeLaFuente:2018uee,Dupuis:2021flq}. For instance, it could be that the first couple orders at large $q$ only receive contributions from the first couple orders in $1/N$.

Finally, we would like to find evidence for other 3d dualities using our method. For instance, if we extend the sub-leading calculation to QED3 with $N$ fermions and nonzero $k$, then we could check other 3d bosonization conjectures such as the duality between the critical $O(2)$ model and QED3 with 1 fermion and $k=1/2$ \footnote{As discussed in \cite{Seiberg:2016gmd}, to properly define QED3 in this case one must add various gravitational Chern-Simons terms, but these do not contribute the local CFT data we are interested in}. We can also consider some QED3 dualities with $N>1$ as discussed in \cite{PhysRevX.7.031051} \footnote{Some evidence from monopole scaling dimensions was found in \cite{Dupuis:2021flq} for the duality between Gross-Neveu QED3 with two fermions and scalar QED3 with two scalars, but it is unclear if the latter theory is in fact a CFT \cite{2019NatPh..15..678Z,Serna:2018tct,Kos:2013tga}, which is a prerequisite for IR duality.}, or dualities with QCD3 for various gauge groups as in \cite{Aharony:2016jvv}. It would also be nice to verify some of the $\mathcal{N}=1$ supersymmetric QED3 dualities such as \cite{Benini:2018umh}.

\section*{Acknowledgments}
We thank Ofer Aharony, Xi Yin, Max Metlitski, Silviu Pufu, Ethan Dyer, Mark Mezei, Chong Wang, and Rufus Boyack for useful conversations, Masataka Watanabe and Rohit Kalloor for collaboration at an early stage of the project, and Ofer Aharony, Subir Sachdev, Senthil Todadri, and Nathan Seiberg for reviewing the manuscript. We also thank the organizers of the Bootstrapping Nature conference in GGI, Florence, during which this project was completed. SMC is supported by the Center for Mathematical Sciences and Applications and the Center for the Fundamental Laws of Nature at Harvard University. WWK is supported by a Discovery Grant from NSERC, a Canada Research Chair, and a grant from the Fondation Courtois.

\appendix

\section{Details of large $N$ calculation}\label{sec:details}

In this appendix we give details for the large $N,k$ calculation of the free energy $F_{q,\kappa}$ in a small temperature expansion for general $q,\kappa$, which gives us the scaling dimension $\Delta_{q,\kappa}$.

\subsection{Leading order}\label{leading}

We start by reviewing the leading order calculation in \cite{Chester:2017vdh}, which is given by plugging in the saddle point values of $A_\mu$ and $\lambda$ \eqref{saddle} into the free energy in \eqref{free} to get
\es{leadF}{
F_{q,\kappa}^{(0)}(\alpha,\mu)=\frac1\beta\tr\log\big[\frac14+\mu-(\nabla_\mu-i\mathcal{A}_\mu)^2\big]-2\kappa q\alpha\,.
}
The eigenvalues of the operator $\left[-(\nabla_\mu-i\mathcal{A}_\mu)^2+\frac14+\mu\right]$ on $S^2\times S^1_\beta$ with magnetic flux $4\pi q$ are $(\omega_n-\alpha)^2+\lambda_j^2$, where $\lambda_j$ are the energies of the modes of the theory quantized on $S^2\times \mathbb{R}$:
\es{spectrum}{
\lambda_j=\sqrt{(j+1/2)^2-q^2+\mu}\,,\quad j\in\{q,\,q+1\,\dots\}\,,
}
with degeneracies $d_j=2j+1$ and $\omega_n=\frac{2\pi n}{\beta}\,,n\in\mathbb{Z}$. We plug this spectrum into \eqref{leadF} and perform the sum to get
\es{free0}{
&F_{q,\kappa}^{(0)}(\alpha,\mu)=-2\kappa q\alpha\\
&\qquad+\beta^{-1}\sum_{j\geq q}d_j\log\left[2\left(\cosh(\beta\lambda_j)-\cosh(\beta\alpha)\right)\right]\,.
}
We then solve the saddle-point equations \eqref{saddleCond} at large $\beta$ to find the set of possible saddle points values of $\alpha$ and $\mu$. For $q=0$, it can be easily checked that $\alpha=\mu=0$ are saddles. For $q>0$, we find the lowest-energy saddle
\es{alphSad}{
\alpha(\kappa)&=-\sgn(\kappa)\big(\lambda_q+\beta^{-1}\log\frac{\xi}{1+\xi}\big)+O(e^{-\beta})\,,\\
 \xi&\equiv\frac{2q|\kappa|}{d_q}\,.
}
We then plug this $\alpha$ into the $\mu$ saddle point equation and zeta-function regularize to get
\es{muSad}{
\sum_{j\geq q}\left(\frac{d_j}{2\lambda_j(\mu)}-1\right)-q+\frac{\xi d_q}{2\lambda_q(\mu)}=0\,.
}
We can numerically extract $\mu$ from this non-linear equation, and we list some example values in Table \ref{leadingTab}. We then plug $\alpha$ and $\mu$ into \eqref{leadF}, take the large $\beta$ limit, and zeta-function regularize to get the leading order scaling dimension
\es{free0final}{
\Delta_{q,\kappa}^{(0)}&=2q|\kappa|\lambda_q+\frac{2q^3+q-6\mu q}{6}\\
&\qquad+\sum_{j\geq q}(d_j\lambda_j-2(j+1/2)^2-\mu+q^2)\,.
}
This expression can be numerically summed for the given $\mu$, and we list some example values in Table \ref{pvtab}. Note that for the special values $\kappa=1$ and $q=1/2$, we find the simple values $\mu=1/4$ and $\Delta_{1/2,1}^{(0)}=1$.

\begin{table}
\centering
\begin{tabular}{c||c|c|c|c|c}
$q$ & 1/2  &1&3/2&2&5/2  \\ 
\hline
$\mu_{\kappa=0}$ &$-.19981 $ &$ -.39782$ &$ -.59546$ &$ -.79294 $& $-.99034$  \\
\hline
$\mu_{\kappa=1}$ &1/4  & $.52026$ & $.79330$ & 1.0672 & 1.34152\\
\hline
\end{tabular}
\caption{Saddle point value $\mu$ for $\kappa=0,1$ for some low $q$.}
\label{leadingTab}
\end{table} 

\subsection{Sub-leading order}\label{subleading}

We will now show the details for the sub-leading calculation. We will follow the $q=1/2$ and $\kappa=1$ discussion in \cite{Chester:2021drl}, except we will generalize to any $q,\kappa$.

\subsubsection{Setup}\label{setup}

 The sub-leading order was written in terms of kernels in \eqref{free1}, which can be written in terms of the Green's function $\langle\phi^i(x)\phi^*_j(x')\rangle=\delta^i_{j}G^q(x,x')$ as
\es{wickCPN}{
{ K}_{q,\text{mat}}^{\mu\nu}(x,x')&=D^\mu G_q(x,x')D^{\nu} G_q(x',x)  \\ 
-G_q(x',x)&D^{\mu}D^{\nu} G_q(x,x') +D^\mu G_q(x',x)D^{\nu} G_q(x,x') \\ 
-G_q(x,x')&D^{\mu}D^{\nu} G_q(x',x)  + 2g^{\mu\nu}\delta(x-x')G_q(x,x)\,,\\
{ K}_{q,\text{CS}}^{\mu\nu}(x,x')&=-{i\kappa\ov 2\pi}\,\delta(x,x') \,\epsilon^{\mu\nu\rho}\partial'_\rho\,,\\
{ K}_q^{\sigma\nu}(x,x')&=G_q(x,x')D^{\nu}G_q(x',x)- G_q(x',x)D^{\nu}G_q(x,x') \,,\\
{ K}_q^{\sigma\sigma}(x,x')&=G_q(x,x')G_q(x',x) \,,
}
where $D^\mu=\partial^\mu-i\mathcal{A}_q^\mu(x)$ and $D^{\nu}=\partial'^{\nu}+i\mathcal{A}_q^{\nu}(x')$. We can then compute the integral over the fluctuations in \eqref{free1} by expanding the fluctuations in Fourier space as 
 \es{aFourier}{
  a(x) &= \mathfrak{a}_{00}^\mathcal{E}(0) \frac{d\tau}{\sqrt{4 \pi \beta}}  + 
  \sum_{n=-\infty}^\infty \sum_{\ell=1}^\infty \sum_{m=-\ell}^\ell 
   \Big[  \mathfrak{a}_{\ell m}^\mathcal{E}(\omega_n) {\cal E}_{n \ell m} (x) \\
& \qquad  + \mathfrak{a}_{\ell m}^\mathcal{B}(\omega_n) {\cal B}_{n \ell m} (x) \Big]  \frac{e^{-i \omega_n \tau}}{\sqrt{\beta}}+ d \lambda(x)\,,\\
      \sigma(x) &=
  \sum_{n=-\infty}^\infty \sum_{\ell=0}^\infty \sum_{m=-\ell}^\ell 
    \mathfrak{b}_{\ell m}(\omega_n) Y_{\ell m} (\theta, \phi) \frac{e^{-i \omega_n \tau}}{\sqrt{\beta}} \,,
 }
where $d\lambda$ are pure gauge modes and ${\cal E}_{n \ell m} (x)$ and ${\cal B}_{n \ell m} (x)$, together with $d\tau/ (4 \pi \beta)$, form an orthonormal basis of polarizations for the one-form $a(x)$:
 \es{EBDef}{
    {\cal E}_{n \ell m}(x) &= \frac{ \ell(\ell+1)  Y_{\ell m}d\tau  -i \omega_n dY_{\ell m}}{\sqrt{\ell(\ell+1)} \sqrt{\omega_n^2 + \ell(\ell+1)}}  \,, \\
     {\cal B}_{n \ell m}(x) &= \frac{*_2 dY_{\ell m}}{\sqrt{\ell (\ell+1)}}  \,,
 }
where $*_2$ is the Hodge dual on $S^2$, and we use the metric
\es{metric}{
ds^2=d\theta^2+\sin^2\theta d\phi^2+d\tau^2\,.
}
We can remove the pure gauge mode contribution to \eqref{free1} by considering $e^{-\beta F_q^{(1)}}/e^{-\beta F_0^{(1)}}$ and using the fact that $F_0^{(1)}=0$. The Fourier transform of the kernels \eqref{wickCPN} for $\ell>0$ is then
\begin{widetext}
\begin{equation}\label{fourierKernscal}
\begin{aligned}
&\bold{K}_{q,\ell}(\omega_n)=\frac{1}{2j+1}\int d^3xd^3x' \sqrt{g}\sqrt{g'} e^{i\omega_n(\tau-\tau')} \sum_{m=-\ell}^{\ell}  \\
& \begin{pmatrix} Y^{\dagger}_{\ell m}(x)K^{\sigma\sigma}_{q}(x,x')  {Y}_{\ell m}(x') & \mathcal{E}^{\dagger}_{\mu,n\ell m}(x)K^{\mu\sigma}_{q}(x,x')  Y_{\ell m}(x')   &   \mathcal{B}^{\dagger}_{\mu,n\ell m}(x)  K^{\mu\sigma}_{q}(x,x')  Y_{\ell m}(x') \\
Y^{\dagger}_{\ell m}(x)K^{\sigma\nu}_{q}(x,x')  \mathcal{E}_{\nu,n\ell m}(x') & \mathcal{E}^{\dagger}_{\mu,n\ell m}(x)K^{\mu\nu}_{q}(x,x')  \mathcal{E}_{\nu,n\ell m}(x')   &   \mathcal{B}^{\dagger}_{\mu,n\ell m}(x)  K^{\mu\nu}_{q}(x,x')  \mathcal{E}_{\nu,n\ell m}(x') \\
Y^{\dagger}_{\ell m}(x)K^{\sigma\nu}_{q}(x,x')  \mathcal{B}_{\nu,n\ell m}(x') &
\mathcal{E}^{\dagger}_{\mu,n\ell m}(x) K^{\mu\nu}_{q}(x,x')  \mathcal{B}_{\nu,n\ell m}(x')   &   \mathcal{B}^{\dagger}_{\mu,n\ell m}(x) K^{\mu\nu}_{q}(x,x')   \mathcal{B}_{\nu,n\ell m}(x') \end{pmatrix}\Bigg\vert_{x'=0}\,,
\end{aligned}
\end{equation}
\end{widetext}
while for $\ell=0$ the only gauge-independent term is the scalar quantity $\bold{K}_{q,0}(\omega_n)\equiv K^{\sigma\sigma}_{q,0}(\omega_n) $. We can then plug \eqref{aFourier}, \eqref{fourierKernscal}, and \eqref{free1} into $e^{-\beta F_q^{(1)}}/e^{-\beta F_0^{(1)}}$, integrate the Fourier modes, and take the large $\beta$ limit to get the sub-leading scaling dimension
\es{freeFinal}{
&\Delta_{q,\kappa}^{(1)}=\int {d\omega\ov 2\pi }\Bigg[\log \le({{\widetilde{ K}}^{\sigma\sigma}_{q, 0}(\omega) \ov {\widetilde{ K}}^{\sigma\sigma}_{0, 0} (\omega) }\ri)\\
&+ \sum_{\ell=1}^\infty  (2\ell+1)\log \det\le(\frac{\widetilde{\bf K}_{q, \ell,\text{mat}}(\omega)+\widetilde{\bf K}_{\ell,\text{CS}}(\omega) }{\widetilde{\bf K}_{0, \ell,\text{mat}}(\omega)+\widetilde{\bf K}_{\ell,\text{CS}}(\omega) }\ri)\Bigg]\,,
}
where the sum over $\omega_n$ has been converted into an integral to exponential precision in $\beta$ and we define the temperature independent kernels $\widetilde{{\bf K}}_{q, \ell}(\omega)$ as
 \es{SmoothApprox}{
  {{\bf K}}_{q, \ell}(\omega_n) = \beta\overline{\bf K}_{q,\ell} \delta_{n0}+\widetilde{{\bf K}}_{q, \ell}(\omega)\big\vert_{\omega=\omega_n}+ O(e^{- \beta}) \,.
}
The linear in $\beta$ terms in the kernels are related to the degeneracy breaking terms in \cite{Chester:2017vdh}, which we will not consider here. 

Our next task is to compute the frequency space kernels. As we see from \eqref{wickCPN}, the Chern-Simons kernel $K_{q,\text{CS}}^{\mu\nu}(x,x')$ is local, so we can compute its frequency space version without doing any integrals to get \cite{Chester:2021drl}
\es{CSfourier2}{
\widetilde{\bold{K}}^{\mathcal{B}\mathcal{E}}_{q,\ell,\text{CS}}(\omega)=\frac{i\kappa}{2\pi}\sqrt{\omega^2+\ell(\ell+1)}\,.
}
The $q=0$ matter kernels can also be easily computed using the closed form expression
\es{freeG}{
G_0(x,x')=\frac{1}{4\pi\sqrt{2\left(\cosh\left(\tau-\tau'\right)-\cos\gamma\right)}}\,,
}
for the $q=0$ Green's function, where $\gamma$ is the angle between the two points on $S^2$
\es{gamma}{
\cos\gamma=\cos\theta\cos\theta'+\sin\theta\sin\theta'\cos\left(\phi-\phi'\right)\,.
}
We then plug this into \eqref{wickCPN}, take the Fourier transform \eqref{fourierKernscal} to compute ${\bf K}_{0,\ell,\text{mat}}(\omega_n)$, and send $\beta\to\infty$ to compute $\widetilde{\bf K}_{0,\ell,\text{mat}}(\omega)$. Some of these kernels are divergent, but as shown in \cite{Dyer:2015zha} we can uniquely regularize using gauge invariance and the $\mu$ saddle point condition to get
\es{q0}{
\widetilde{K}^{\sigma\sigma}_{0,\ell}(\omega)&=\left|\frac{\Gamma\left((\ell+1+i\omega)/2\right)}{4\Gamma\left((\ell+2+i\omega)/2\right)}\right|^2\,,\\
\widetilde{K}^{\mathcal{B}\mathcal{B}}_{0,\ell}(\omega)&=\frac{\omega^2+\ell^2}{2}\widetilde{K}^{\sigma\sigma}_{\ell-1}(\omega)\,,\\
\widetilde{K}^{\mathcal{E}\mathcal{E}}_{0,\ell}(\omega)&=\frac{\omega^2+\ell(\ell+1)}{2}\widetilde{K}^{\sigma\sigma}_{\ell}(\omega)\,,
}
while the other $q=0$ kernels vanish.

\subsubsection{Matter kernels}\label{matter}

The $q\neq0$ matter kernels are much more complicated. We start by writing down the $q\neq0$ Green's function \cite{Chester:2017vdh}
\es{GreenDef2}{
G_q(x,x')&=e^{\alpha(\tau-\tau')} \left(G(x,x')+ \hat G(x,x')  \right)\,,\\
G(x,x')&\equiv\sum_{j,m}{Y_{q,jm}(\theta,\phi)Y^*_{q,jm}(\theta',\phi')} \int\frac{d\omega}{2\pi}\frac{e^{-i\omega(\tau-\tau')}}{\omega^2+\lambda_j^2}\,,\\
\hat G(x,x')&\equiv  \frac{ q|\kappa| e^{\sgn(\kappa)\lambda_q(\tau-\tau')} }{(2q+1)\lambda_q} \sum_{m}{Y_{q,qm}(\theta,\phi)Y^*_{q,qm}(\theta',\phi')}  \,,
}
where $G(x,x')$ is the same as the $\kappa=0$ $CP$-invariant Green's function computed in \cite{Dyer:2015zha}, while $\hat G(x,x')$ is a new $CP$-violating contribution due to the nontrivial $\alpha$ that only appears because $k\neq0$. Both of these are written in terms of monopole spherical harmonics as defined in \cite{Wu:1976ge,Wu:1977qk}.  

All the terms that appear in the matter kernels \eqref{wickCPN} either contain two Green's functions $G_{q}(x,x')$ and $G_{q}(x',x)$, or $G_{q}(x,x)\delta(x,x')$, so the overall phase $e^{\alpha(\tau-\tau')}$ can be neglected as long as we replace the covariant derivative $D_\mu$ by $\hat D_\mu\equiv D_\mu\vert_{(\nabla_\tau-\alpha)\to\nabla_\tau}$. The resulting terms can then be divided into three categories. Firstly, there are terms that come only from $ \hat G(x,x')$ that are independent of $\tau$, and so only contribute to the linear in $\beta$ term $\overline{\bf K}_{q,j} $ defined in \eqref{SmoothApprox} that do not contribute to the scaling dimension. Secondly, there are terms coming only from pairs of $  G(x,x')$, and so preserve $CP$. After taking $\beta\to\infty$ these terms are identical to the $k=0$ calculation in \cite{Dyer:2015zha}, except that the saddle point value $\mu$ will differ for $k\neq0$. As discussed in \cite{Dyer:2015zha}, there are divergences in the calculation, which can be regularized exactly as they did. The non-zero $CP$ terms were computed in a non-gauge-invariant basis in \cite{Dyer:2015zha}, which can be related to our gauge invariant basis as
\es{CPnonzero2}{
\widetilde{K}^{\sigma\sigma}_{q,\ell}(\omega)&=D_{q,\ell}(\omega)\,,\quad
\widetilde{K}^{\mathcal{B}\mathcal{B}}_{q,\ell}(\omega)={K}^{{B}{B}}_{q,\ell}(\omega)\,,\\
 \widetilde{K}^{\mathcal{E}\mathcal{E}}_{q,\ell}(\omega)&=\left(1+\frac{\omega^2}{\ell(\ell+1)}\right){K}^{{\tau}{\tau}}_{q,\ell}(\omega)\,,\\
  \widetilde{K}^{\mathcal{B}\sigma}_{q,\ell}(\omega)&={H}^{ B}_{q,\ell}(\omega)\,.\\
}
We then simply take the explicit expressions for the RHS from \cite{Dyer:2015zha}. For instance, for $\widetilde{K}^{\sigma\sigma}_{q,\ell}(\omega)$ we have
\es{dql}{
&\widetilde{K}^{\sigma\sigma}_{q,\ell}(\omega)= \frac{8 \pi^2}{2 \ell + 1} \sum_{j', j''=q}^{\infty}\Bigg[  \frac{ \lambda_{j'}  + \lambda_{ j''} }
{
  \lambda_{j'}  \lambda_{j''} ( \omega^2 + ( \lambda_{j'} +  \lambda_{j''})^2)
}\\
&   \frac{(2 \ell + 1) (2 j' + 1) (2 j'' + 1)}{64 \pi^3} 
   \begin{pmatrix}
    \ell & j' & j'' \\
    0 & -q & q
   \end{pmatrix}^2\Bigg]\,,
}
where the 3-j symbols makes the $j'$ sum finite, but there is still an infinite sum over $j''$. The other kernels in \cite{Dyer:2015zha} are written in terms of a quantity 
 \es{CqDef}{
  C_q \equiv 
    \frac{1}{4 \pi} \left[ \sum_{j' = q}^{\infty} \left( \frac{j' +1/2}{\sqrt{(j' + 1/2)^2 + \mu_q^2 - q^2}} - 1 \right) - q \right] \,.
 }
which was zero for the $k=0$ theory, but for general $k$ we see from saddle point equation \eqref{muSad} that it is
\es{Cqus}{
C_q=-\frac{q|\kappa|}{4\pi\lambda_q}\,.
}
In the attached \texttt{Mathematica} file we give explicit expressions for all these $CP$ kernels \footnote{We correct some typos in the expressions as written in \cite{Dyer:2015zha}.}, which are all written as infinite sums like \eqref{dql}.

 The third and last set of terms in $\eqref{wickCPN}$ come from pairs of $ G(x,x')$ and $\hat G(x',x)$ or just a single $\hat G(x',x)$ and so violate $CP$. This means that all of the terms in the kernel matrix \eqref{fourierKernscal} will be nonzero, which includes both the kernels in \eqref{CPnonzero2} as well as the new entirely $CP$-violating terms
 \es{newKerns}{
   \widetilde{K}^{\mathcal{E}\sigma}_{q,\ell}(\omega)\,,\qquad  \widetilde{K}^{\mathcal{E}\mathcal{B}}_{q,\ell}(\omega)\,.
 }
The nice thing about these kernels is that their calculation is entirely finite, so we don't need to deal with any new regularization subtleties relative to \cite{Dyer:2015zha}. Another nice thing is that all the intermediate sums are finite, so there will be no infinite sums like in \eqref{dql}. A new difficulty, though, is that the position space kernels now include both terms that are exponentially damped in $\tau$, as well as terms that are independent of $\tau$. If we perform the $\tau$ integral in \eqref{fourierKernscal} and then send $\beta\to\infty$, the former terms contribute to $\widetilde{\bold K}_{q,j,NCP}(\omega)$ while the latter terms are linear in $\beta$ and so contribute to $\overline{\bold K}_{q,j}(\omega_n)$, as defined in \eqref{SmoothApprox}. If we send $\beta\to\infty$ before we compute the $\tau$ integral, then these linear in $\beta$ terms will appear as delta functions in $\omega$, so we define $\widetilde{\bold K}_{q,j,NCP}(\omega)$ by first sending $\beta\to\infty$, then performing the $\tau$ integral, then throwing out any $\delta(\omega)$ that appear. 

For $q=1/2$ and $\kappa=1$, the $NCP$ kernels were computed by just plugging the Green's function $G_q(x,x')$ into \eqref{wickCPN} and \eqref{fourierKernscal} and performing the integrals explicitly, which was possible because the Green's function takes a simplified form in this case. For general $q$ and $\kappa$, however, $G_q(x,x')$ is written in terms of two monopole harmonics, so that \eqref{fourierKernscal} is a double integral of four monopole harmonics and two vector harmonics. Thankfully, it was shown in Appendix B of \cite{Dyer:2015zha} how these integrals can be performed in terms of sums of 3-j symbols, which is in fact how \eqref{dql} was derived. For instance, to compute $\widetilde{ K}^{\sigma\sigma}_{q,j,NCP}(\omega)$ we take the position space expression in \eqref{wickCPN} and plug in the part of the Green's function \eqref{fourierKernscal} that will give the $NCP$ term to get
\es{DNCP1}{
&K^{\sigma\sigma}_{q,NCP}(x,x')=G(x,x') e^{-\sgn(\kappa)\lambda_q(\tau-\tau')}\hat G(x',x) \\
&+G(x',x) e^{\sgn(\kappa)\lambda_q(\tau-\tau')}\hat G(x,x') \,.
}
We then plug this into \eqref{fourierKernscal}, take $\beta\to\infty$, and apply the identities in Appendix B of \cite{Dyer:2015zha} to get
\es{DNCP2}{
&\widetilde{ K}^{\sigma\sigma}_{q,\ell,NCP}(\omega)=\sum_{j'=\max\{q,\ell-q\}}^{\ell+q} \Big[ \Big(
\begin{array}{ccc}
 \ell & j' & q \\
 0 & -q & q \\
\end{array}
\Big)^2 \\
& \qquad\qquad\qquad \qquad\qquad\frac{(2\ell+1)q |\kappa|(\omega^2+\lambda_{j'}^2-\lambda^2_q)}{2\pi \lambda_q((\omega^2+\lambda_q^2+\lambda_{j'}^2)^2-4\lambda_q^2\lambda_{j'}^2)}\Big]\\
&=\sum_{j'=\max\{q,\ell-q\}}^{\ell+q}\frac{(2\ell+1)q |\kappa|(\omega^2+\lambda_{j'}^2-\lambda^2_q)}{2\pi(2j'+1) \lambda_q((\omega^2+\lambda_q^2+\lambda_{j'}^2)^2-4\lambda_q^2\lambda_{j'}^2)}\,,
}
which is a finite sum for all $\ell,q$. Note that we have consistently ignored any delta functions that appear in intermediate steps, following the discussion below \eqref{SmoothApprox}.  We can similarly compute the entirely $CP$-violating term
\es{HtNCP2}{
&\widetilde{ K}^{\mathcal{E}\tau}_{q,\ell,NCP}(\omega)=\frac{(2\ell+1)q\kappa}{\pi}\sqrt{1+\frac{\omega^2}{\ell(\ell+1)}}\\
&\times\sum_{j'=\max{q,\ell-q}}^{\ell+q}\frac{(\lambda_{j'}^2-\lambda^2_q)}{ (2j'+1)((\omega^2+\lambda_q^2+\lambda_{j'}^2)^2-4\lambda_q^2\lambda_{j'}^2)} \,,
}
which is also a finite sum. The other $NCP$ kernels can be computed similarly and result in uglier expressions that we write explicitly in the attached \texttt{Mathematica} notebook. We can then check these general $\kappa,q$ expression against the $\kappa=1\,, q=1/2$ expression in 4.38 of \cite{Chester:2021drl}.

\subsubsection{Numerical sum/integral with asymptotic}\label{num}

Finally, we plug the $q=0$ and $q\neq0$ Fourier space kernels into \eqref{freeFinal} and take the determinant to get the sub-leading scaling dimension, which we write as
\es{subScal}{
\Delta_{q,\kappa}^{(1)}=\frac12\int\frac{d\omega}{2\pi}\sum_{\ell=0}^\infty(2\ell+1)L^{q,\kappa}_{\ell}(\omega)\,.
}
Note that $\sgn\kappa$ factors in all matter kernels cancel, so the expression only depends on $|\kappa|$. We can compute the large $\omega$ and $\ell$ asymptotic of $L_{\ell}(\omega)$ by a variation of the calculation in Appendix C of \cite{Dyer:2015zha}, which we briefly sketch here. We start by writing the differential equation that defines the Green's function: 
\es{GreenDef}{
\left[-\left(\nabla_\mu-i\mathcal{A}_\mu\right)^2+\frac14+\mu\right]G_q(x,x')=\delta(x-x')\,.
}
Previously we gave the exact solution of this equation in \eqref{GreenDef2}, but we can also solve it perturbatively in $t\equiv\sinh^2\tau/2$ and $s\equiv \sin^2\theta/2$. At each order the solution will have an integration constant, that can be fixed by explicit comparison to the exact solution in \eqref{GreenDef2}. This perturbative expression for the Green's function will thus be identical to the $\kappa=0$ expression in Appendix C of \cite{Dyer:2015zha}, except the integration constants will differ due to the new second term in \eqref{GreenDef2}. As shown in that paper, the large $s,t$ expansion of $G_q(x,x')$ can then be plugged into the Fourier kernels \eqref{fourierKernscal} to get the large $\omega$ and $\ell$ asymptotic for the kernels in \eqref{CPnonzero2}, including both the $CP$ preserving and violating contributions. For the other kernels in \eqref{newKerns}, we note that they are finite sums like \eqref{HtNCP2}, so they can be easily expanded. We then combine these expressions to get the asymptotic
\es{largeL}{
L^{q,\kappa}_\ell(\omega)=\left(\frac{8\mu}{\omega^2+(\ell+1/2)^2}\right)\left(\frac{1}{1+\frac{64\kappa^2}{\pi^2}}\right)+\dots\,,
} 
where the leading terms is linear divergent and we give higher terms in the attached \texttt{Mathematica} file. This linear divergence is regularized by counterterms, which are equivalent to the zeta function regularization
\es{zetaL}{
&\left(\frac{1}{1+\frac{64\kappa^2}{\pi^2}}\right)\int\frac{d\omega}{4\pi}\sum_{\ell=0}^\infty\frac{8\mu(2\ell+1)}{\omega^2+(\ell+1/2)^2}\\
&\qquad\qquad\qquad=4\mu\frac{\zeta(0,1/2)}{1+\frac{64\kappa^2}{\pi^2}}=0\,.
}
We can then subtract \eqref{zetaL} from \eqref{subScal} to get the expression
\es{subScalReg}{
&\Delta_{q,\kappa}^{(1)}=\frac12\int\frac{d\omega}{2\pi}\sum_{\ell=0}^\infty(2\ell+1)\Big[L^{q,\kappa}_{\ell}(\omega)\\
&\qquad\qquad-\left(\frac{8\mu}{\omega^2+(\ell+1/2)^2}\right)\left(\frac{1}{1+\frac{64\kappa^2}{\pi^2}}\right)\Big]\,.
}
Even after the linear divergence has been regularized, there is still a potential logarithmic divergence in the integral in \eqref{subScalReg}. This logarithmic divergence cancels as long as we regularize this integral consistent with conformal symmetry, such as by using the symmetric cutoff
\es{cutoff}{
(\ell+1/2)^2+\omega^2<\Lambda^2\,.
}
To obtain good precision, we first evaluate \eqref{subScalReg} numerically in a region $(\ell+1/2)^{2}+\omega^{2}<(\Lambda')^{2}$;  then in the region $(\Lambda')^2 < (\ell+1/2)^{2}+\omega^{2} < \Lambda^2$ we replace $L_\ell^{q,\kappa}(\omega)$ in \eqref{subScalReg} by the sub-leading convergent terms in the asymptotic expansion \eqref{largeL}, as given in the attached \texttt{Mathematica} notebook up to order $O(1/\left[(\ell + \frac 12)^2 + \omega^2 \right]^{7/2})$), and then evaluate the integral analytically as $\Lambda \to \infty$.  We notice that the result converges very rapidly as we increase $\Lambda'$ for general $q,\kappa$, which was used to compute the values shown in Table \ref{ktab}. We give some additional values for $\kappa=2$ in Table \ref{k2tab}.

\begin{table}
\centering
\begin{tabular}{c||c|c|c}
$q$ & $ \Delta_{q,2}^{(0)} $ &  $\Delta_{q,2}^{(1)}$ &  $N=1$   \\ 
\hline
\hline
$1/2$ & 2.1101 & $-0.3218$ & 1.788 \\
$1$ & 5.5781 & $-0.6906$ & 4.888  \\
$3/2$ & 10 & $-1.124$ & 8.876 \\
$2$ & 15.2027 & $-1.617$ & 13.59 \\
$5/2$ & 21.0831  & $-2.162$ & 18.92  \\
\hline
\end{tabular}
\caption{Scaling dimensions $\Delta_{q,2}=N\Delta_{q,2}^{(0)}+\Delta_{q,2}^{(1)}+O(1/N)$ for charge $q$  monopole operators in QED3 with $N$ scalars and $k/N=2$ in a large $N,k$ expansion extrapolated to $N=k/2=1$. The dual theory is conjectured~\cite{Seiberg:2016gmd} to be a gauged Dirac fermion with Chern-Simons level $k=-3/2$. 
}
\label{k2tab}
\end{table}

\section{Counting fermions}\label{sec:count}

In this appendix we count the number of lowest dimension operators of charge $q$ in a 3d free fermion theory using the character formulae in \cite{Dolan:2005wy}. The $S^2\times S$ partition function is defined for a general 3d CFT with $U(1)$ flavor symmetry as 
\es{Zdef}{
Z(s,f,x)=\sum_{\cO_{\Delta,j,q}} s^\Delta f^{2q} x^j\,,
}
which runs over all operators $\cO_{\Delta,j,q}$ in the CFT with dimensions $\Delta$, spin $j$, and $U(1)$ charge $q$. Note that these operators include descendents, so to count primary operators we should use the character formulae for a general $\Delta>j+1$ operator
\es{genOp}{
\chi_{\Delta,j}(s,x)=\frac{s^\Delta\chi_j} {(1-s)(1-sx)(1-s/x)}\,,
}
a conserved $\Delta=j+1$ current 
\es{curr}{
\chi_{j+1,j}(s,x)=\frac{s^{j+1}(\chi_j-s\chi_{j-1})}{(1-s)(1-sx)(1-s/x)}\,,
}
and a $\Delta=1$ free fermion
\es{freeFerm}{
\chi_{1,1/2}(s,x)=\frac{s(\sqrt{x}+1/\sqrt{x})}{(1-sx)(1-s/x)}\,,
}
where we define the spin $j$ character
\es{chi}{
\chi_j=\frac{x^{j+1/2-x^{-j-1/2}}}{\sqrt{x}-1/\sqrt{x}}\,.
}
The partition function for the free fermion theory is 
\es{Zfree}{
Z(s,f,x)=\exp\Big[\sum_{n=1}^\infty\frac{f^n+f^{-n}}{n}(-1)^{n+1}\chi_{1,1/2}(s^n,x^n)\Big]\,,
}
which can be expanded in terms of the above characters (multiplied by suitable powers of $f$) get the following lowest dimension operators $(\Delta,j)$ for the lowest several $q$:
\es{listOps}{
q&=1/2:\qquad \{(1,1/2)\}\\
q&=1:\qquad \{(2,0)\}\\
q&=3/2:\qquad \{(4,3/2)\}\\
q&=2:\qquad \{(6,0)\,,\quad (6,2)\}\\
q&=5/2:\qquad \{(8,3/2)\}\\
q&=3:\qquad \{(10,0)\}\\
q&=7/2:\qquad \{(13,5/2)\}\\
q&=4:\qquad \{(16,0)\,,\quad\ (16,2)\,,\quad (16,4)\}\\
q&=9/2:\qquad \{(19,3/2)\,,\quad (19,5/2)\,,\quad (19,9/2)\}\\
q&=10:\qquad \{(22,2)\,,\quad (22,4)\}\\
q&=11/2:\qquad \{(25,5/2)\}\\
q&=6:\qquad \{(28,0)\}\,.
}
These values were used to write the free fermion dimensions in Table \ref{ktab}. Note that the only $q$ values
that have a unique scalar operator are the $q$ listed in \eqref{uniqueq}.

\onecolumngrid
\vspace{1in}
\twocolumngrid

\bibliographystyle{ssg}
\bibliography{sat}

\end{document}